\documentclass[twocolumn,reprint,showpacs, superscriptaddress, amsmath,amssymb, aps, prl]{revtex4-1}
\usepackage{graphicx}
\usepackage{amsmath}
\usepackage{epsfig}
\usepackage{array}
\usepackage{multirow}
\usepackage{color}
\usepackage{dcolumn}
\usepackage{bm}
\usepackage{float}
\usepackage[
   bookmarks=true,
   colorlinks=true,
   hyperindex=true,
   citecolor=blue,
   linkcolor=blue,
]{hyperref}
\begin{document}

\title{Revealing Energy Dependence of Quantum Defects via Two Heteronuclear Atoms in an Optical Tweezer}

\author{Kunpeng Wang}
\affiliation{State Key Laboratory of Magnetic Resonance and Atomic and Molecular Physics, Wuhan Institute of Physics and Mathematics, Chinese Academy of Sciences - Wuhan National Laboratory for Optoelectronics, Wuhan 430071, China}
\affiliation{Center for Cold Atom Physics, Chinese Academy of Sciences, Wuhan 430071, China}
\affiliation{School of Physics, University of Chinese Academy of Sciences, Beijing 100049, China}

\author{Xiaodong He}
\email{hexd@wipm.ac.cn}
\affiliation{State Key Laboratory of Magnetic Resonance and Atomic and Molecular Physics, Wuhan Institute of Physics and Mathematics, Chinese Academy of Sciences - Wuhan National Laboratory for Optoelectronics, Wuhan 430071, China}
\affiliation{Center for Cold Atom Physics, Chinese Academy of Sciences, Wuhan 430071, China}

\author{Xiang Gao}
\email{xiang.gao@tuwien.ac.at}
\affiliation{Institute for Theoretical Physics, Vienna University of Technology, A-1040 Vienna, Austria}
\affiliation{Beijing Computational Science Research Center, Beijing 100193, China}

\author{Ruijun Guo}
\affiliation{State Key Laboratory of Magnetic Resonance and Atomic and Molecular Physics, Wuhan Institute of Physics and Mathematics, Chinese Academy of Sciences - Wuhan National Laboratory for Optoelectronics, Wuhan 430071, China}
\affiliation{Center for Cold Atom Physics, Chinese Academy of Sciences, Wuhan 430071, China}
\affiliation{School of Physics, University of Chinese Academy of Sciences, Beijing 100049, China}

\author{Peng Xu}
\affiliation{State Key Laboratory of Magnetic Resonance and Atomic and Molecular Physics, Wuhan Institute of Physics and Mathematics, Chinese Academy of Sciences - Wuhan National Laboratory for Optoelectronics, Wuhan 430071, China}
\affiliation{Center for Cold Atom Physics, Chinese Academy of Sciences, Wuhan 430071, China}

\author{Jun Zhuang}
\affiliation{State Key Laboratory of Magnetic Resonance and Atomic and Molecular Physics, Wuhan Institute of Physics and Mathematics, Chinese Academy of Sciences - Wuhan National Laboratory for Optoelectronics, Wuhan 430071, China}
\affiliation{Center for Cold Atom Physics, Chinese Academy of Sciences, Wuhan 430071, China}
\affiliation{School of Physics, University of Chinese Academy of Sciences, Beijing 100049, China}

\author{Runbing Li}
\affiliation{State Key Laboratory of Magnetic Resonance and Atomic and Molecular Physics, Wuhan Institute of Physics and Mathematics, Chinese Academy of Sciences - Wuhan National Laboratory for Optoelectronics, Wuhan 430071, China}
\affiliation{Center for Cold Atom Physics, Chinese Academy of Sciences, Wuhan 430071, China}

\author{Min Liu}
\affiliation{State Key Laboratory of Magnetic Resonance and Atomic and Molecular Physics, Wuhan Institute of Physics and Mathematics, Chinese Academy of Sciences - Wuhan National Laboratory for Optoelectronics, Wuhan 430071, China}
\affiliation{Center for Cold Atom Physics, Chinese Academy of Sciences, Wuhan 430071, China}

\author{Jin Wang}
\affiliation{State Key Laboratory of Magnetic Resonance and Atomic and Molecular Physics, Wuhan Institute of Physics and Mathematics, Chinese Academy of Sciences - Wuhan National Laboratory for Optoelectronics, Wuhan 430071, China}
\affiliation{Center for Cold Atom Physics, Chinese Academy of Sciences, Wuhan 430071, China}

\author{Jiaming Li}
\affiliation{Department of Physics and Center for Atomic and Molecular Nanosciences, Tsinghua University, Beijing 100084, China}
\affiliation{Key Laboratory for Laser Plasmas (Ministry of Education), and Department of Physics and Astronomy, Shanghai Jiao Tong University, Shanghai 200240, China}
\affiliation{Collaborative Innovation Center of Quantum Matter, Beijing 100084, China}

\author{Mingsheng Zhan}
\email{mszhan@wipm.ac.cn}
\affiliation{State Key Laboratory of Magnetic Resonance and Atomic and Molecular Physics, Wuhan Institute of Physics and Mathematics, Chinese Academy of Sciences - Wuhan National Laboratory for Optoelectronics, Wuhan 430071, China}
\affiliation{Center for Cold Atom Physics, Chinese Academy of Sciences, Wuhan 430071, China}
\date{\today}

\begin{abstract}
As a physically motivated and computationally simple model for cold atomic and molecular collisions, the multichannel quantum defect theory (MQDT) with frame transformation (FT) formalism provides an analytical treatment of scattering resonances in an arbitrary partial wave between alkali-metal atoms, leading to the experimental observation of $p-$ and $d-$wave resonances. However, the inconsistency of quantum defects for describing scattering resonances shows up when compared with experiments. Here, with two heteronuclear atoms in the ground state of an optical tweezer, the energy dependence of quantum defects is obviously revealed by comparing the measured s-wave scattering length with the prediction of MQDT-FT. By dividing the quantum defects into energy sensitive and insensitive categories, the inconsistency is ultimately removed while retaining the analytic structure of MQDT-FT. This study represents a significant improvement in the analytical MQDT-FT and demonstrates that a clean two-particle system is valuable to the test of collisional physics.
\end{abstract}


\maketitle
The tremendous progress in laser cooling, trapping and manipulating ultracold matters allows us to unveil a host of unique phenomena in quantum mechanical nature and thus attracts wide interest, reaching far into other fields, such as condensed matter and few- and many-body physics, beyond the atomic and molecular physics \cite{Bloch08RMP,Chin2010,Baranov2012,Greene2017}. To determine the scattering properties and reveal the corresponding prospects of precise control of cold gaseous matter, a detailed understanding of the involved collision processes in the constituents is crucial. This can be obtained via a powerful yet analytical theoretical tool: the multichannel quantum defect theory (MQDT) \cite{Seaton1983,Fano1975,Greene1985,Aymar1996}. This theory was initially developed by Seaton for atomic system with a long-range Coulomb interaction \cite{Greene2017} and has been applied successfully for various atomic \cite{Fano1975,Aymar1996,Lee1973}, and molecular collisions \cite{Greene1985,Fano1970,Jungen1998,GaoX2010}, as well as cold atom and cold molecule collisions ~\cite{Burke1998,Idziaszek2010,Croft2012}.

In particular, for the cold atom collisions, by replacing the numerical solutions for the long range potential with the analytic ones for the dominant Van der Waals interaction ($-C_6/R^6$) between pairs of neutral atoms \cite{Gao1998a}, the scattering processes, including shape and Feshbach resonances, can be described by a set of three parameters: two eigenchannel quantum defects $\mu_{\alpha}$ (singlet and triplet quantum defects $\mu_{s}$ and $\mu_{t}$) that describe the scattering phase shifts at small inter-particle separations $R$ and the dispersion coefficient $C_6$ that describes the potentials at large $R$ \cite{Gao2005,Hanna2009,Gao2009,Gao2011,Ruzic2013}. The predictive power of the analytic MQDT-FT, especially in search of resonances in $p$- and $d$-wave, has been proven experimentally \cite{Pwave,YouLidwave,Yao2019}. However, the derived triplet quantum defects deviate from the ones offered by numerical coupled-channel calculations based on the full knowledge of the molecular potentials of the collisional pairs, e.g. for the case of $^{87}$Rb-$^{85}$Rb systems \cite{Pwave,YouLidwave,Strauss2010} and $^{6}$Li-$^{133}$Cs system \cite{Pires2014}.

To uncover more evidences of the underlying physics giving rise to the discrepancies between the MQDT-FT and experiments involved, complementary to the ones caught in studies of $^{87}$Rb-$^{85}$Rb magnetic Feshbach resonances, here we test the MQDT-FT by applying it to an inherently simplistic and seemingly straightforward problem, that is, determination of $s$-wave scattering lengths of open channels that has not yet been examined. Experimentally, such a scattering length is measured by utilizing an ultracold pair of $^{87}$Rb-$^{85}$Rb atoms with nearly perfect wavefunction overlap in a single optical tweezer, instead of the conventional bulk samples of cold atoms. To this end, we measure the microwave (MW) transition shifts for different scattering channels, and then deduce the associated scattering length from the analytical solution under the pseudopotential approximation. When comparing our measurements with predictions of MQDT-FT without adjusting quantum defects, discrepancies show up as encountered in experiments with Feshbach resonances. We find that the  discrepancies originate from the energy dependence of quantum defects due to the contributions from $-C_8/R^8$ and $-C_{10}/R^{10}$ interactions outside the reaction zone, which are typically ignored in previous MQDT-FT. When we introduce two triplet quantum defects according to the energy sensitivity, the discrepancies can be resolved. Thus our explanation reveals the energy-dependence of quantum defects and can be extended to other complex dynamical processes and systems.

In the following, the scattering channels are defined by the internal states of $^{87}$Rb ($q1$) and $^{85}$Rb ($q2$), where the $q1$ ($q2$) denotes the quantum numbers of Zeeman sublevels $|F,m_F\rangle$ of $^{87}$Rb ($^{85}$Rb). A scattering channel is conveniently labeled by specifying the set of quantum numbers \{$q1$; $q2$\} afterward (only include the $m_F$ quantum number for brevity).

\begin{figure}
	\centering
	\includegraphics[width=\linewidth]{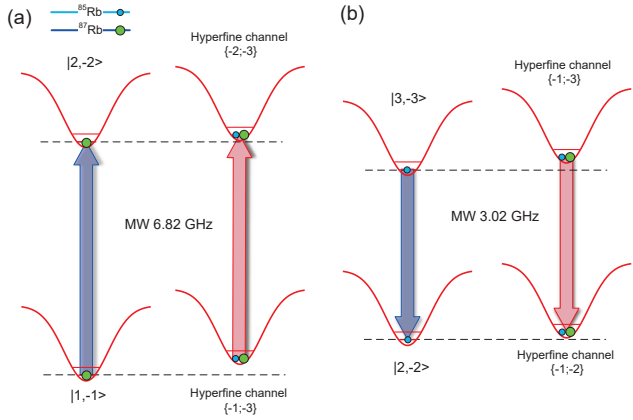}
	\caption{Experimental scheme of the measurement of collisional shifts between $^{87}$Rb and $^{85}$Rb atoms in the motional ground states. (a) For the MW transition of $^{87}$Rb, single atom transition starts from $|1,-1\rangle$ to $|2,-2\rangle$ and the involved channels in two atom collisions are \{-1;-3\} and \{-2;-3\}. (b) For $^{85}$Rb, single atom transition starts from $|3,-3\rangle$ to $|2,-2\rangle$ and the involved hyperfine channels in collisions are \{-1;-3\} and \{-1;-2\}. }
	\label{setup}
\end{figure}

\begin{figure}
	\centering
	\includegraphics[width=\linewidth]{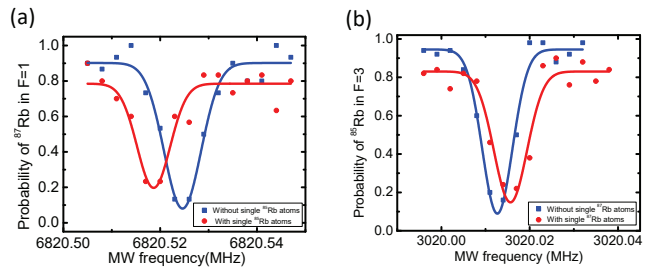}
	\caption{Collisional shifts of single $^{87}$Rb and $^{85}$Rb atoms. (a) The measured MW resonant spectral of $^{87}$Rb from $|1,-1\rangle$ to $|2,-2\rangle$ transition in the presence (red dots) and in the absence (blue squares) of $^{85}$Rb in the $|3,-3\rangle$  state.  (b) The measured MW resonant spectral of $^{85}$Rb from $|3,-3\rangle$ to $|2,-2\rangle$ transition in the presence (red dots) and in the absence (blue squares) of $^{87}$Rb in the $|1,-1\rangle$ state. }
	\label{2bodyMW}
\end{figure}

\begin{figure}
	\centering
	\includegraphics[width=\linewidth]{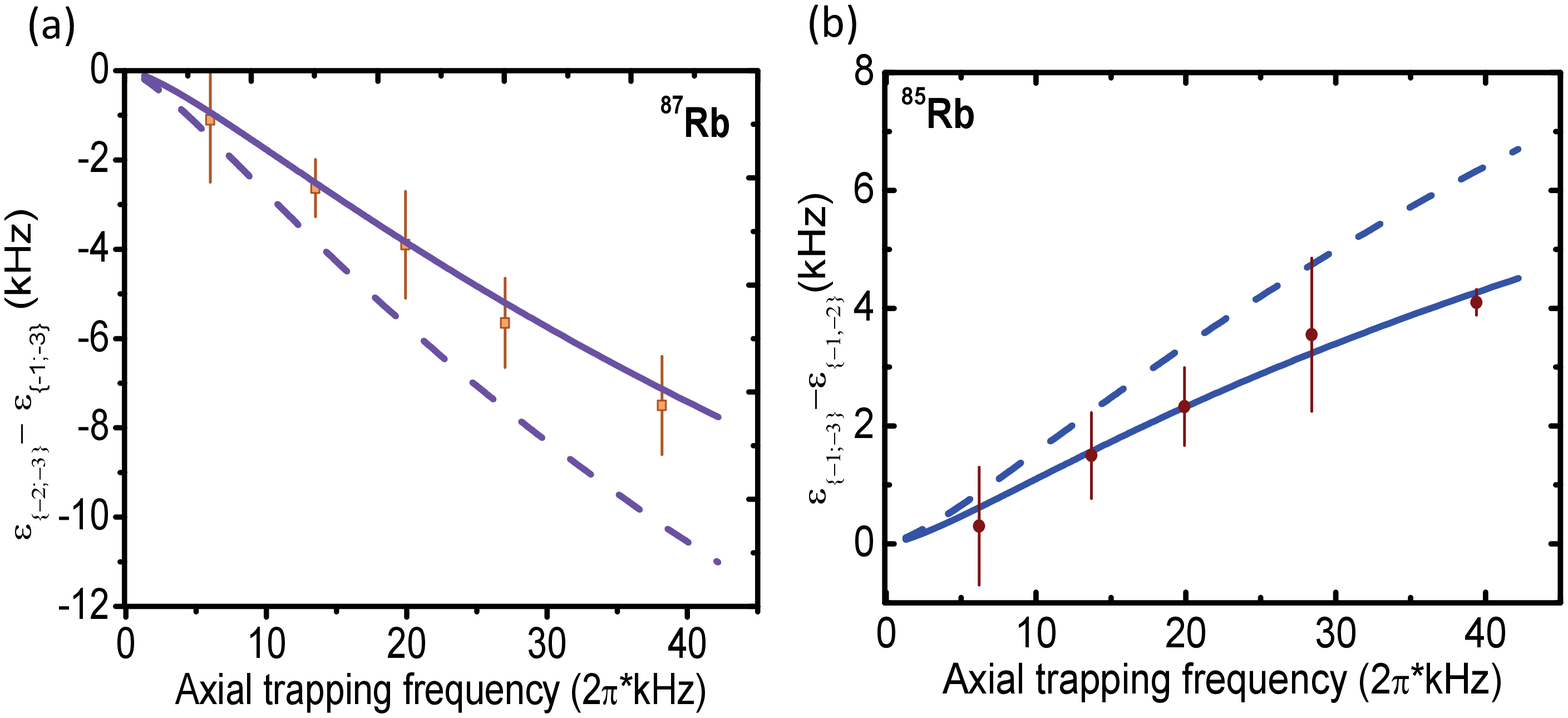}
	\caption{The dependencies of collisional shifts experienced by $^{87}$Rb atoms (see the left panel (a)) and $^{85}$Rb atoms (b) on the axial oscillating frequencies. For (a) and (b), each dot is an average of 100-150 times of measurements. The accompanying error bars are statistic standard deviation for the average. The solid lines are fits to the Eq.(1) using the method of least-squares minimization and the dashed lines are the predictions of the shifts by adopting the values of quantum defects determined from precision spectroscopic measurements~\cite{Strauss2010} as input parameters in the analytical MQDT-FT. See the main text for more details.}
	\label{2bodyinteraction}
\end{figure}

To experimentally deduce the scattering length, as shown in Fig. \ref{setup}, we measure the resonant frequency difference of the MW transitions between two-atom collisions and the single-atom transitions. This scheme is free of the mean field shifts that are normally encountered in a bulk sample of cold atoms.
When interrogating the $^{87}$Rb atom, single-atom transition starts from $|F,m_F\rangle \equiv |1,-1\rangle$ to $|2,-2\rangle$, and the corresponding collisions start from \{-1; -3\} channel to \{-2; -3\}. When interrogating $^{85}$Rb, single-atom transition starts from $|3,-3\rangle$ to $|2,-2\rangle$, and the corresponding collisions start from \{-1; -3\} channel to \{-1; -2\}. The main experimental details on the preparation of the ultracold sample of two atoms were described in ref. \cite{Wang2019}. The corresponding resonant frequencies are measured by implementing the conventional Rabi spectroscopy, which is an excellent probe for interacting cold atoms \cite{Campbell2006,Goban2019,Franchi2017}.
The spectroscopy on the $^{87}$Rb ($^{85}$Rb) atom is implemented with a 120 $\mu$s (near $\pi$) MW pulse, as shown in Fig. \ref{2bodyMW}(a) (Fig. \ref{2bodyMW}(b)). For these measurements, the trap oscillation frequencies of single atoms are about 165 kHz and 27 kHz in the radial and axial directions respectively. As shown in Fig. \ref{2bodyMW}(a), by comparing with that of a single $^{87}$Rb, the resonant frequency is found to be reduced by $5.6\pm1.0$ kHz as colliding with a $^{85}$Rb populated in the state of $|3,-3\rangle$, while the resonant frequency of $^{85}$Rb atom is increased by $3.5\pm1.3$ kHz due to colliding with a $^{87}$Rb atom in the state of $|1,-1\rangle$, as shown in Fig. \ref{2bodyMW}(b). To efficiently extract the scattering lengths, we measure the frequency shifts of $^{87}$Rb and $^{85}$Rb atoms as functions of the trapping frequencies, as shown in the Fig. \ref{2bodyinteraction}(a) and (b) respectively. We note that, the large measurement errors in transition frequencies are mainly caused by the fluctuations of magnetic field, since the Zeeman states involved are magnetic sensitive.

For the rubidium isotopes in a Gaussian optical tweezer with a wavelength of 852 nm, the difference in trap frequencies of the two atoms is only about 1$\%$ due to the slight difference in masses and the almost same trap depths (difference of $10^{-4}$). The trapping frequencies of the two atoms are thus approximately equal so that the center of mass and relative motion are separable. We can thus straightforwardly adopt the analytical solutions for the dynamics of two trapped interacting ultracold atoms under the energy-independent delta pseudopotential approximation~\cite{Idziaszek2005,Idziaszek2006} to evaluate the interaction energy. The pseudopotential approximation is valid provided that the van der Waals length scale $\beta_6=(2 \mu C_6/\hbar^2)^{1/4}$ is less than the radial harmonic-oscillator width $d_r=\sqrt{\hbar/(\mu\omega_{r})}$, where $\mu$ is the reduced mass of the atom pair, $\omega_{r}$ is the angular trapping oscillation frequency in the radial direction, e.g. see refs. ~\cite{Blume2002,Bolda2002}.
For the measurements in Fig. \ref{2bodyinteraction}(a) and (b), the smallest $d_r\approx$ 595 $a_0$, and is much larger than the $\beta_6\approx$165.1 $a_0$, given the $C_6\approx$ 4710 a.u. (atomic units), where $a_0$ is the Bohr radius. Therefore all of the data adopted fulfill the aforementioned validity of pseudopotential approximation.

The ratio of the trapping frequencies in the radial and axial directions is  $\eta\approx$ 6. The derived relationship between collisional energy $\varepsilon_{\{q1;q2\}}$ and $s$-wave scattering length $a_{\{q1;q2\}}$ of scattering channel \{q1;q2\} is given by
\begin{equation}
\label{Vr}
-\frac{\sqrt{\pi}}{a_{\{q1;q2\}}}=\mathcal{F}(-\varepsilon_{\{q1;q2\}}/2),
\end{equation}
where $\mathcal{F}(-\varepsilon_{\{q1;q2\}}/2)$ denotes an implicit function of $\varepsilon_{\{q1;q2\}}$~\cite{Idziaszek2005,Idziaszek2006}, and  $\mathcal{F}(x)=-2\sqrt{\pi}\Gamma(x)/\Gamma(x-1/2)+\sqrt{\pi}\Gamma(x)/\Gamma(x+1/2)\sum_{m=1}^{n-1}F(1,x;x+1/2;\exp(i 2\pi m/n))$, where F(a,b;c;x) denotes the hypergeometric function, and n is a positive integer, here $n=\eta \approx 6$. Thus, Eq.(1) relates the $a_{\{q1;q2\}}$ to the collisional energy $\varepsilon_{\{q1;q2\}}$.

Given one of the scattering length of the associated channels, e.g. \{-2;-3\} channel~\cite{Bloch2001}, the actual collisional energy $\varepsilon_{\{-2;-3\}}$ can be calculated by using the Eq.(\ref{Vr}), and thus the $\varepsilon_{\{-1;-3\}}$ of the channel to be measured can be deduced from the measured frequency shifts. Ultimately, the scattering length $a_{\{-1;-3\}}$ can be extracted by fitting the data points in Fig. \ref{2bodyinteraction}(a) to the Eq.(\ref{Vr}).

The associated scattering channels in the Fig. \ref{2bodyinteraction}(a) are \{-2;-3\} and \{-1;-3\}, therefore the measured frequencies amounts to the difference between two interaction energies $\epsilon_{\{-2;-3\}}$ and $\epsilon_{\{-1;-3\}}$. Given the measured value of $a_{\{-2;-3\}}=213(7) \, a_0$ in the previous work \cite{Bloch2001}, the data in Fig. \ref{2bodyinteraction}(a) are fitted to Eq. (\ref{Vr}) via least-squares minimization yielding  $a_{\{-1;-3\}}=3.2(5)\times 10^2 \, a_0$, as shown by the solid line. Then we turn to the collisional shifts of $^{85}$Rb atoms, as shown in Fig. \ref{2bodyinteraction}(b), which associate with the differences between two other interaction energies, and for scattering channels $\epsilon_{\{-1;-2\}}$ and $\epsilon_{\{-1;-3\}}$, respectively. Given the calculated $a_{\{-1;-2\}}=229.4 \, a_0$ via coupled channel method \cite{Strauss2010}, the other measured value of $a_{\{-1;-3\}}$  is similarly extracted from the same fitting function as $3.0(5)\times 10^2 \, a_0$, as the solid line shown in Fig. \ref{2bodyinteraction}(b). The average value of $a_{\{-1;-3\}}=3.1(5)\times 10^2 \, a_0$. This measurement matches the coupled channel calculations using the MOLSCAT package \cite{Hutson2018} with the precise interaction potentials derived from high precision molecular spectroscopy \cite{Strauss2010}.

Subsequently following the analytical MQDT-FT approach \cite{Gao2005} and adopting the \{$\mu_s$, $\mu_t$\} $\approx$ \{0.7253, 0.1822\} determined from precision spectroscopic measurements \cite{Strauss2010}, we found that the calculated scattering length $a_{\{-1;-3\}}$  is obviously overestimated, as shown in Table \ref{table1}. To intuitively illustrate such a discrepancy, the calculated corresponding dependence of collisional shifts seen by $^{87}$Rb on the axial trapping frequencies is plotted as the dashed lines shown in Fig. \ref{2bodyinteraction}(a), as well as the similar behavior encountered in $^{85}$Rb shown in Fig. \ref{2bodyinteraction}(b). Additionally, we also calculate the scattering length of channel \{-1;-2\}, given the same input parameters of  $\mu_s$ and $\mu_t$, similar discrepancies show up, as shown in Table \ref{table1}. Interestingly, using the $\mu_t$  determined by another cold atom experiments \cite{Pwave}, the calculated scattering lengths for these two elastic channels are both underestimated, as shown in Table \ref{table1}.

\begin{table*}
	\caption{List of scattering lengths for \{-1;-3\} and \{-1;-2\} channels. The resulting scattering lengths are divided into two groups by adopting two different sets of input parameters of $\{\mu_s,\mu_t\}$ for the MQDT-FT, from up to down, the values of $\{\mu_s,\mu_t\}$ are respectively adopted from ref.~\cite{Strauss2010} and ref.~\cite{Pwave}. And each group is compared to both the measured result and coupled-channel calculations. See the main text for more details. The scattering lengths are in units of $a_0$. }
	\begin{tabular}{c|cc|ccc}
		\hline
		\hline
		Channel&$\mu_s$&$\mu_t$&$a_{MQDT-FT}$&$a_{exp}$&$a_{CC}$\\
		\hline			
		\{-1;-3\} & $0.7253$ & 0.1822 & 420.2&3.1(5)$\times10^2$ & 314.8 \\
		\{-1;-2\} & $0.7253$ & 0.1822 &242.5& &229.4 \\		
		\hline
		\{-1;-3\} & $0.7253$ &0.2045& 277.4&3.1(5)$\times10^2$ & 314.8  \\			
		\{-1;-2\} & $0.7253$ &0.2045& 207.7& & 229.4 \\
		\hline
		\hline
	\end{tabular}\label{table1}
\end{table*}

\begin{table*}
	\caption{List of improved scattering lengths for \{-1;-3\} and \{-1;-2\} channels. The resulting scattering lengths are calculated by adopting the improved set of parameters $\{\mu_s^{EI},\mu_t^{EI},\mu_t^{ES}\}$  and are compared with both the measured result and coupled-channel calculations. See the main text and Supplemental Material for more details.}
	\begin{tabular}{c|ccc|ccc}
		\hline
		\hline
		Channel&$\mu_s^{EI}$&$\mu_t^{EI}$& $\mu_t^{ES}$&$a_{MQDT-FT}$&$a_{exp}$&$a_{CC}$\\
		\hline
		\{-1;-3\} & $0.7253$ & 0.1822 &$0.1984$ &315.0&3.1(5)$\times10^2$ &314.8 \\			
		\{-1;-2\} & $0.7253$ & 0.1822 &$0.1984$  &234.3& & 229.4 \\
		\hline
		\hline
	\end{tabular}\label{table2}
\end{table*}


In the analytical framework of the MQDT-FT with only $C_6$ type of long-range potential, one need to choose a very large reaction zone where the $C_8$, $C_{10}$ interactions are negligible. But after carrying out a careful check for the Rb$_2$ interaction potentials \cite{Strauss2010}, we find out that the long-range $C_8$ and $C_{10}$ potentials play a crucial role in determining the values of scattering lengths in the region $20 \, a_0 \leq R \leq 45 \, a_0$, where the effective momentum for channels with high dissociative thresholds may be comparable with the contributions of $C_8$ and $C_{10}$ interactions, leading that the $\mu_{\alpha}$ of these eigenchannels \cite{Supp} are much more sensitive to the channel energy than other eigenchannels. Based on the phase amplitude method or the Wentzel-Kramers-Brillouin (WKB) approximation, the eigen quantum defect $\mu_{\alpha}$ closely related with the inverse (approximately) of the local effective momentum of each channel within the reaction zone \cite{Burke1998}. Then one would expect a strong perturbation to $\mu_{\alpha}$ if there exists closed channels with the zero of their local effective momentum $k(R)$ within the extended reaction zone. Therefore, after the extension of the reaction zone, the original eigen quantum defect $\mu_{\alpha}$ will change by  $\delta\mu_{\alpha}$, which represents the corrections due to the high-order long-range potentials, and may strongly depend on the channel energies.

Based on the conventional MQDT, the scattering state wavefunctions can be expressed as superpositions of eigenchannel wavefunctions $\Psi_\alpha$ characterizing the detailed dynamics within the reaction zone with a common short-range phase shift, which relates to the eigenchannel quantum defect $\mu_\alpha$ by $\pi\mu_\alpha$. It was customarily expected that the $\mu_\alpha$ is almost independent of the channel energies, e.g. see Refs.~\cite{Gao2005,Pwave,YouLidwave}.
However, from the above arguments, to retain the analytic structure of MQDT-FT and meanwhile remove the differences between theoretical and experimental results on the scattering lengths, we propose to divide the eigenchannels into two categories: the energy insensitive (EI) one and the energy sensitive (ES) one.

For our interested \{-1; -3\} elastic scattering process, the singlet (0, 0; 4, -4) and triplet (1, 0; 4, -4) eigenchannels are almost correlated with the two fragmentation channels with the two lowest dissociative thresholds, the zero point of their $k(R)$ are larger than $45\, a_0$ and the $C_8$ and $C_{10}$ potentials will generally influence their $k(R)$ within the reaction zone by less than 10\%. We classify them into the EI category, the corresponding values of  $\mu_{s}^{EI}$ and  $\mu_{t}^{EI}$ remain 0.7253 and 0.1822, respectively. On the other hand, the remaining two triplet (1,-1; 3,-3) and (1,-1; 4,-3) eigenchannels are correlated with the two highest dissociative thresholds, the zero point of their $k(R)$ are at about $41 \, a_0$ and the influence of $C_8$ and $C_{10}$ potential can be larger than 50\%. We then classify them into the ES category, where a slightly different value $\mu_{t}^{ES} \approx 0.1984$ is assigned. Similarly, we use the same classification standard and the related  $\mu_{\alpha}$ values for the \{-1; -2\} scattering process with total M=-3. As shown in Table \ref{table2}, the resulting scattering lengths for both \{-1; -3\} and \{-1; -2\} channels are significantly improved and in good agreement with our experimental measurement and the coupled-channel calculations.

To confirm the plausibility of our proposal to introduce the energy dependence of quantum defects, we also carried out some numerical studies for calculating the eigenchannel quantum defects from first-principle, in a similar spirit of what we did in atomic system \cite{GaoX2016}. The preliminary results have confirmed the correctness of our classification of $\mu_{t}$ into the energy sensitive and insensitive ones, hence this study represents a significant improvement in the analytical MQDT-FT.

Extending the improved MQDT-FT scenario to the analysis of Feshbach resonances in ultracold ensembles is straightforward \cite{Pwave,YouLidwave,Yao2019,Pires2014}. In general, the scattering length measurement technique introduced here can be used to test the theoretical models of collisions beyond the neutral atoms \cite{Campbell2006,Goban2019,Franchi2017}, such as atom-molecule and molecule-molecule scattering processes. Furthermore, the present work sets the stage for experiments with a deterministically prepared two-particle system in which clean collisional dynamics and Feshbach resonances can be further explored.

\begin{acknowledgments}
We acknowledge E. Tiemann for his coupled-channel calculation results on the scattering lengths. We also thank L. You for fruitful discussions. This work was supported by the National Key Research and Development Program of China under Grant No. 2017YFA0304501, No. 2016YFA0302104, No. 2016YFA0302800, and No. 2016YFA0302002, the National Natural Science Foundation of China under Grant No. 11774389, No. 11774023 and No. U1530401, the Strategic Priority Research Program of the Chinese Academy of Sciences under Grant No. XDB21010100 and the Youth Innovation Promotion Association CAS No. 2019325.
\end{acknowledgments}

\bibliographystyle{apsrev4-1} 

\newpage

\appendix
\setcounter{equation}{0}

\begin{appendix}

\section{Supplemental Material}

\author{Kunpeng Wang}
\affiliation{State Key Laboratory of Magnetic Resonance and Atomic and Molecular Physics, Wuhan Institute of Physics and Mathematics, Chinese Academy of Sciences - Wuhan National Laboratory for Optoelectronics, Wuhan 430071, China}
\affiliation{Center for Cold Atom Physics, Chinese Academy of Sciences, Wuhan 430071, China}
\affiliation{School of Physics, University of Chinese Academy of Sciences, Beijing 100049, China}

\author{Xiaodong He}
\email{hexd@wipm.ac.cn}
\affiliation{State Key Laboratory of Magnetic Resonance and Atomic and Molecular Physics, Wuhan Institute of Physics and Mathematics, Chinese Academy of Sciences - Wuhan National Laboratory for Optoelectronics, Wuhan 430071, China}
\affiliation{Center for Cold Atom Physics, Chinese Academy of Sciences, Wuhan 430071, China}

\author{Xiang Gao}
\email{xiang.gao@tuwien.ac.at}
\affiliation{Institute for Theoretical Physics, Vienna University of Technology, A-1040 Vienna, Austria}
\affiliation{Beijing Computational Science Research Center, Beijing 100193, China}

\author{Ruijun Guo}
\affiliation{State Key Laboratory of Magnetic Resonance and Atomic and Molecular Physics, Wuhan Institute of Physics and Mathematics, Chinese Academy of Sciences - Wuhan National Laboratory for Optoelectronics, Wuhan 430071, China}
\affiliation{Center for Cold Atom Physics, Chinese Academy of Sciences, Wuhan 430071, China}
\affiliation{School of Physics, University of Chinese Academy of Sciences, Beijing 100049, China}

\author{Peng Xu}
\affiliation{State Key Laboratory of Magnetic Resonance and Atomic and Molecular Physics, Wuhan Institute of Physics and Mathematics, Chinese Academy of Sciences - Wuhan National Laboratory for Optoelectronics, Wuhan 430071, China}
\affiliation{Center for Cold Atom Physics, Chinese Academy of Sciences, Wuhan 430071, China}

\author{Jun Zhuang}
\affiliation{State Key Laboratory of Magnetic Resonance and Atomic and Molecular Physics, Wuhan Institute of Physics and Mathematics, Chinese Academy of Sciences - Wuhan National Laboratory for Optoelectronics, Wuhan 430071, China}
\affiliation{Center for Cold Atom Physics, Chinese Academy of Sciences, Wuhan 430071, China}
\affiliation{School of Physics, University of Chinese Academy of Sciences, Beijing 100049, China}

\author{Runbing Li}
\affiliation{State Key Laboratory of Magnetic Resonance and Atomic and Molecular Physics, Wuhan Institute of Physics and Mathematics, Chinese Academy of Sciences - Wuhan National Laboratory for Optoelectronics, Wuhan 430071, China}
\affiliation{Center for Cold Atom Physics, Chinese Academy of Sciences, Wuhan 430071, China}

\author{Min Liu}
\affiliation{State Key Laboratory of Magnetic Resonance and Atomic and Molecular Physics, Wuhan Institute of Physics and Mathematics, Chinese Academy of Sciences - Wuhan National Laboratory for Optoelectronics, Wuhan 430071, China}
\affiliation{Center for Cold Atom Physics, Chinese Academy of Sciences, Wuhan 430071, China}

\author{Jin Wang}
\affiliation{State Key Laboratory of Magnetic Resonance and Atomic and Molecular Physics, Wuhan Institute of Physics and Mathematics, Chinese Academy of Sciences - Wuhan National Laboratory for Optoelectronics, Wuhan 430071, China}
\affiliation{Center for Cold Atom Physics, Chinese Academy of Sciences, Wuhan 430071, China}

\author{Jiaming Li}
\affiliation{Department of Physics and Center for Atomic and Molecular Nanosciences, Tsinghua University, Beijing 100084, China}
\affiliation{Key Laboratory for Laser Plasmas (Ministry of Education), and Department of Physics and Astronomy, Shanghai Jiao Tong University, Shanghai 200240, China}
\affiliation{Collaborative Innovation Center of Quantum Matter, Beijing 100084, China}

\author{Mingsheng Zhan}
\email{mszhan@wipm.ac.cn}
\affiliation{State Key Laboratory of Magnetic Resonance and Atomic and Molecular Physics, Wuhan Institute of Physics and Mathematics, Chinese Academy of Sciences - Wuhan National Laboratory for Optoelectronics, Wuhan 430071, China}
\affiliation{Center for Cold Atom Physics, Chinese Academy of Sciences, Wuhan 430071, China}

\date{\today}

\maketitle

\subsection{Details of Multichannel Quantum Defect Theory Calculation}
For two colliding alkali-metal atoms in their ground states, the eigenchannels can be simply denoted as (S,M$_s$; I,M$_I$), where S=s$_1$+s$_2$ and I=i$_1$+i$_2$ are the total electronic spin and nuclear spin of the two atoms respectively, and M$_s$ and M$_I$ are the corresponding projection of them on the interatomic axis respectively. Since s$_1$=s$_2$=1/2, we have singlet and triplet eigenchannels, the corresponding quantum defects are described by $\mu_s$ and $\mu_t$, respectively, denoted as $\mu_\alpha=\{\mu_s,\mu_t\}$. When the colliding atoms outside the reaction zone, the eigenchannel wavefunctions $\Psi_\alpha$ as can also be expressed as superpositions of fragmentation channels wavefunctions $\Phi_i$. For present elastic scattering channel of \{-1;-3\} with total M = -4, there are four eigenchannels involved, namely, (0,0; 4,-4), (1,0; 4,-4), (1,-1; 3,-3) and (1,-1; 4,-3), which relate to the fragmentation channel \{-1;-3\} by the superposition coefficients $U_{i\alpha}$, reflecting the detailed interaction dynamics. Under the FT approximation~\cite{Fano1970-2,Lee1973-2,Gao1996-2}, $U_{i\alpha}$ can be simply calculated by the geometric recoupling coefficients, i.e.
\begin{equation}
\label{eq2}
U_{i\alpha}=\langle F^i_1M^i_{F_1}F_2^iM_{F_2}^i|S^\alpha M_S^\alpha I^\alpha M_I^\alpha\rangle
\end{equation}

It should be mentioned that although the FT are broadly used in atomic~\cite{Lee1973-2,Lee1975-2}, molecular~\cite{Fano1970-2,Jungen1998-2,GaoX2010-2} and cold atom collision problems~\cite{Gao1996-2,Pwave-2,YouLidwave-2} with great success, there exists some differences with the ab-initio calculation results~\cite{Lee1973-2,Greene1985-2}.

From the eigenchannel parameters of $\mu_\alpha$ and $U_{i\alpha}$ , the short-range scattering matrix $K^c_{ij}$ is expressed as~\cite{Gao2005-2}:
\begin{equation}
K^c_{ij}=\Sigma_\alpha U_{i\alpha} \tan(\pi\mu_\alpha+\pi/8)U_{j\alpha}.
\end{equation}
For our interested scattering channel \{-1; -3\}, the effective one open channel short-range scattering matrix $K^c_{eff}$ is obtained by projecting out the other three closed channels~\cite{Burke1998-2,Gao2005-2}:
\begin{equation}
K^c_{eff}=K^c_{oo}+K^c_{oc}(\chi^c-K^c_{cc})^{-1}K^c_{co}.
\end{equation}
where $\chi^c$(E) is defined through the large-R asymptotic behaviors of the negative energy solutions of $f^c$ and $g^c$ for the $C_6$ long-range interactions~\cite{Gao2005-2}, which are -0.8155366, 2.5661999, and 2.5668389 for the $\{F=2,m_F=-2; F=2,m_F=-2\}$, $\{F=2,m_F=-2; F=3,m_F=-2\}$ and $\{F=2,m_F=-1; F=3,m_F=-3\}$ closed channels, respectively. Using the analytical solutions of $C_6$ interaction potential, the scattering length can be expressed as~\cite{Gao2005-2}:
\begin{equation}
a_{\{q1;q2\}}=\frac{2^{2/3}\pi}{[\Gamma(1/4)]^2}\frac{K^c_{eff}+\tan(\pi/8)}{K^c_{eff}-\tan(\pi/8)}(2\mu C_6/\hbar^2)^{1/4}.
\end{equation}
Then if the $\{\mu_s,\mu_t\}$ and the $C_6$ are given, the scattering length can be obtained by applying the above calculation. The resulting scattering lengths are summarized in the Table I and II by adopting different set of $\{\mu_s,\mu_t\}$ with various conditions.

\end{appendix}

\end{document}